\documentclass[final,5p,times,twocolumn]{elsarticle}
 
\journal{Physics Letters B}

\usepackage{etoolbox}
\usepackage{slashed,physics,graphicx}
\usepackage{cuted}
\usepackage{xcolor}

\patchcmd{\emailauthor}{(#2)}{}{}{}

\begin{document}

\begin{frontmatter}

\title{Freeze-In Dark Matter within the Seesaw mechanism}

\author[unibo,infn,rwth]{Michele Lucente}
\ead{michele.lucente@unibo.it}

\address[unibo]{Dipartimento di Fisica e Astronomia, Universit\`a di Bologna, via Irnerio 46, 40126 Bologna, Italy}
            
\address[infn]{INFN, Sezione di Bologna, viale Berti Pichat 6/2, 40127, Bologna, Italy}

\address[rwth]{Institute for Theoretical Particle Physics and Cosmology (TTK), RWTH Aachen University, D-52056 Aachen, Germany}

\date{\today}

\begin{abstract}
We show that the minimal Type-I Seesaw mechanism can successfully account for the observed dark matter abundance in the form of a keV sterile neutrino. This population can be produced by the decay of the  heavier neutral leptons, with masses above the electroweak mass scale, while they are in thermal equilibrium in the early Universe (freeze-in). Moreover, the implementation of the relevant phenomenological constraints (relic abundance, indirect detection and structure formation) on this model automatically selects a region of the parameter space featuring an approximate lepton number symmetry.
\end{abstract}


\end{frontmatter}

\section{Introduction}
The nature of dark matter (DM) and the mechanism at the origin of neutrino masses (and lepton flavour mixing) are among the most pressing  questions in particle physics and cosmology. A minimal and natural extension of the Standard Model (SM) of particle physics is the Type-I Seesaw mechanism~\cite{Minkowski:1977sc,GellMann:1980vs,Mohapatra:1979ia,Yanagida:1980xy,Schechter:1980gr,Schechter:1981cv}, in which the field content of the SM is extended by  including neutrino fields with right-handed chirality (RHN). Being gauge singlets, RHN can have Majorana mass terms whose origin is not related to the SM Higgs mechanism, thus leaving the RHN mass scale as a phenomenological free parameter; for the same reason, the number of RHN fields is not fixed by the requirement of anomaly cancellation, although this is not generally true in the context of gauge extensions of the SM, where the number of RHN has to match the number of SM generations ($n=3$) to ensure anomaly cancellation (cf. e.g.~\cite{Mohapatra:1998rq}). We also notice that, in the context of the minimal SM, there exists only one global symmetry which is anomaly free, given by the difference between the baryon number $B$ and lepton number $L$, i.e. $B-L$.

Sterile neutrinos with masses at the keV scale can be suitable DM candidates (c.f.~\cite{Boyarsky:2018tvu} for a review), subject to a number of phenomenological constraints concerning their current abundance~\cite{Aghanim:2018eyx}, lifetime~\cite{Pal:1981rm} and impact on structure formation (notably as inferred by observations of the Lyman-$\alpha$ forest~\cite{Tremaine:1979we,Boyarsky:2008ju,Hui:1996fh,Gnedin:2001wg,Weinberg:2003eg,Boyarsky:2008xj}): the first two mainly constrain the mixing  $\theta$ between active and sterile neutrinos, while the latter strongly depends on the DM production mechanism.
These particles are unavoidably produced in the early Universe as long as a non-zero mixing with the active neutrinos is present (Dodelson-Widrow mechanism, DW)~\cite{Dodelson:1993je}. For masses below 28.8 keV (95\% C.L.), this production mechanism results in a DM population with a too large free-streaming length (classified as warm dark matter, WDM), incompatible with the observed large scale structure of the Universe~\cite{Baur:2017stq}. For larger masses, the constraints on $\theta$ derived from the non-observation of the sterile neutrino DM decay products~\cite{Boyarsky:2005us,Boyarsky:2006fg,Watson:2006qb,Yuksel:2007xh,Boyarsky:2007ge,Loewenstein:2008yi,RiemerSorensen:2009jp,Mirabal:2010an,Essig:2013goa,Horiuchi:2013noa,Riemer-Sorensen:2014yda,Tamura:2014mta,Ng:2015gfa,Riemer-Sorensen:2015kqa,Neronov:2016wdd,Perez:2016tcq,Dessert:2018qih,Boyarsky:2020hqb,Ng:2019gch,Roach:2019ctw,DeRomeri:2020wng,Foster:2021ngm} prevent the DW production to account for the observed DM relic density.
The conversion of active into sterile neutrinos can be resonantly enhanced if the primordial plasma features a sizeable lepton asymmetry (Shi-Fuller mechanism, SF)~\cite{Shi:1998km}; the effect is twofold: on one hand, smaller values of $\theta$ are required to account for the observed DM relic density, thus evading the bounds from indirect detection. On the other hand, the DM free streaming is reduced with respect to DW, thus resulting in a colder DM population and relaxing structure formation bounds. These ideas have been coherently implemented in the so-called $\nu$MSM model~\cite{Asaka:2005an,Asaka:2005pn}, which is a phenomenologically motivated implementation of the Type-I Seesaw mechanism with 3 RHN: one of them at the keV scale acting as DM candidate, while the other two have strongly degenerate masses at the GeV scale and, after electroweak symmetry breaking (EWSB), produce a lepton asymmetry which acts as a background for the resonant DM production via SF mechanism. Moreover, the same heavy sterile neutrinos can as well account for the observed Baryon Asymmetry of the Universe via leptogenesis before EWSB~\cite{Canetti:2012kh,Ghiglieri:2020ulj}.

In this Letter, we point out that an alternative implementation of the minimal Type-I Seesaw mechanism exists, which is able to account for the observed DM relic density\footnote{Sterile neutrino DM has been proposed as well in models with an extended gauge or scalar sector, see e.g.~\cite{Khalil:2008kp,Abada:2014zra,Seto:2020udg,DeRomeri:2020wng}.}. The DM candidate is still a keV sterile neutrino, but the other sterile neutrinos have a much larger mass scale, above the electroweak one. They produce the correct DM abundance while in thermal equilibrium by decaying into a keV sterile neutrino and  a Higgs boson: this production mechanism has been firstly pointed out in~\cite{Abada:2014zra} in the framework of a minimal Inverse Seesaw realisation~\cite{Abada:2014vea}.

\section{The model}
We extend the SM field content with the addition of 3 RHN fields $N_I$, $I=1,2,3$, leading to the most general renormalizable Lagrangian:
\begin{multline}\label{eq:seesaw_lag}
	\mathcal{L} = \mathcal{L}_\mathrm{SM} + i \overline{N_I}\slashed{\partial}N_I 
	-\left(\frac{1}{2}\overline{N_I^c} M_{IJ} N_J + F_{\alpha I} \overline{\ell_L^\alpha} \tilde{\phi} N_I + h.c.\right),
\end{multline}
where $\ell_L^\alpha$ is the left-handed $SU(2)_L$ lepton doublet of flavour $\alpha=e,\mu,\tau$, $M_{IJ}$ is a symmetric matrix of Majorana mass terms, $F_{\alpha I}$ are dimensionless Yukawa couplings and  $N^c = i \gamma^2\gamma^0 \overline{N}^T$, $\tilde{\phi}=i\sigma^2 \phi^*$.
After EWSB, the Yukawa couplings generate a Dirac mass term $m_D = v F$, where $v$ is the Higgs vacuum expectation value (vev, taking the value $v_0=174$ GeV at zero temperature), and Eq.~(\ref{eq:seesaw_lag}) results in a non-vanishing Majorana mass matrix $m_\nu$ for the light active neutrinos; in the Seesaw limit, $|M_{IJ}| \gg v|F_{\alpha I}|$, it reads $m_\nu \simeq - v^2 F^T M^{-1} F$, while at leading order in the expansion parameter $\Theta=v F/M$ the heavier mass eigenstates have masses coinciding with the eigenvalues of the matrix $M$. Without loss of generality, it is always possible to chose a basis in which the matrix $M$ is real and diagonal, in which case the mixing elements coupling the active lepton flavour $\alpha$ to the heavy mass eigenstate $I$ are given by $U_{\alpha I}^* = m_D^{\alpha I}/M_I$.
In order to ensure the agreement of the model with experimental neutrino mixing parameters, it is convenient to parametrise the Yukawa couplings using the Casas-Ibarra (CI) parametrisation~\cite{Casas:2001sr}:
\begin{equation}\label{eq:CI}
	m_D = - i\ U_\mathrm{PMNS}^* \sqrt{\hat{m}}\ R\ \sqrt{M},
\end{equation}
where $\hat{m}$ is a diagonal matrix containing the mass eigenvalues of the light (mostly active) neutrinos at zero temperature, $U_\mathrm{PMNS}$ is the unitary PMNS mixing matrix~\cite{Zyla:2020zbs} and $R$ is an orthogonal matrix parametrised by 3 complex angles $\omega_{ij}$,
\begin{equation}
	R = V_{23} V_{13} V_{12},
\end{equation}
with
\begin{equation}
V_{12} = \left(
	\begin{array}{ccc}
	\cos\omega_{12} & \sin\omega_{12}& 0 \\
	-\sin\omega_{12} & \cos\omega_{12} & 0 \\
	0 & 0 & 1
	\end{array}
	\right),
\end{equation}
and analogous definitions for $V_{23},V_{13}$.

\subsection{Model parameters}
Low energy neutrino oscillation data  fix the value of $U_\mathrm{PMNS}$ and (squared) mass differences for $\hat{m}$ in Eq.~(\ref{eq:CI}). In order to be a viable DM candidate, a sterile neutrino must have a mass at the keV scale~\cite{Boyarsky:2018tvu}: we will thus explore realisations with $M_1$ in the range $[1,100]$ keV. The heavier mass eigenstates must lie above the electroweak scale for the freeze-in production mechanism proposed in~\cite{Abada:2014zra} to take place, but not too far from it; for definiteness, we will chose them to be almost degenerate and fix the values $M_{2,3}=300$ GeV. The large hierarchy of values between $M_1$ and $M_{2,3}$ may seem a fine-tuned scenario, however we will show that this can be the natural result of an approximate lepton number symmetry. Most importantly, this symmetry-motivated scenario gets automatically realised once we impose astrophysical and cosmological constraints to the model. 

\subsection{Phenomenological constraints}
\subsubsection{Cosmology and astrophysics}
There are three broad categories of constraints that apply to the model under discussion. The first requirement is to not overclose the Universe, by producing a larger DM abundance than the observed one, $\Omega^\mathrm{obs}_\mathrm{DM} h^2 = 0.12$~\cite{Aghanim:2018eyx} (where $h$ is the dimensionless Hubble parameter, $H_0 = 100\ h\ \textrm{km s}^{-1}\ \textrm{Mpc}^{-1}$). In the absence of a sizeable lepton asymmetry, there are two production mechanisms to consider: the first one is the well known DW, effective at temperatures $T\approx 150$ MeV and resulting in a relic density
\begin{equation}\label{eq:DW}
	\Omega_\mathrm{DM}^{DW} h^2 = 0.11\cdot 10^5 \frac{M_1}{\mathrm{keV}} \sum_\alpha C_\alpha (M_1) \left|\left(U_\mathrm{PMNS}^* \sqrt{\frac{\hat{m}}{\mathrm{eV}}} R\right)_{\alpha 1}\right|^2,
\end{equation}
where $C_\alpha$ are dimensionless parameters of order unity, whose exact value depends on the lightest sterile neutrino mass $M_1$~\cite{Asaka:2006nq}. It is evident from Eq.~(\ref{eq:DW}) that the requirement $\Omega_\mathrm{DM}^{DW} h^2\leq 0.12$ imposes a strong upper bound on the elements of the first column of $R$, and/or an upper bound on the light active neutrino masses (see~\cite{Boyarsky:2006jm} for an extended discussion on this second point).
The second production mechanism, i.e. freeze-in production from the decay of heavy neutrinos in thermal equilibrium, has never been included (to the best of our knowledge) in previous studies concerning the minimal Type-I Seesaw mechanism. It is mainly effective at temperatures $T\approx M_{2,3}$, but the related couplings are different from zero only after EWSB; its resulting relic abundance can be estimated as, 
\begin{equation}\label{eq:FI_relic}
	\Omega_\mathrm{DM}^{FI} h^2 = 2.16\cdot 10^{22} \sum_{J=2,3} \left|\left(R^\dagger \frac{\hat{m}}{v} R\right)_{1J} \right|^2 g_{J} \left(1-\frac{m_h^2}{M_J^2}\right)^2\varepsilon\left(M_J\right),
\end{equation}
with $m_h$ the Higgs boson mass and $g_J$  the internal degrees of freedom of the heavy neutrinos. The function $\varepsilon(M_J)$ was defined in~\cite{Abada:2014zra} (following~\cite{Hall:2009bx}), and takes into account the evolution of the Higgs vev $v$ and the suppression of the decay rate in DM due to the electroweak symmetry restoration for decays happening at temperatures larger than the EWSB scale (notice that since $m_\nu \propto v^2$ the ratio $m_\nu/v$ vanishes for $v=0$). The factorisation in Eq.~(\ref{eq:FI_relic}) neglects the evolution of the Higgs mass with $v$ for temperatures approximately in the range 160 to 130 GeV, where the sharp electroweak crossover takes place~\cite{DOnofrio:2014rug}, thus slightly underestimating the available phase space in the decay and thus the final DM abundance.
  For sub-eV active neutrinos, we have $(\hat{m}/v_0)^2 \lesssim 10^{-23}$, giving the DM abundance in the correct ballpark, provided $R$ is of order unity.
  We notice that, although other decay channels mediated by the SM gauge bosons are potentially relevant at zero temperature, thermal effects strongly suppress the production of sterile neutrinos from gauge vector boson decays at electroweak temperatures~\cite{Lello:2016rvl}; we thus neglect these decay channels as they give a negligible contribution to the final abundance.

The second constraint comes from requiring the DM stability over cosmological timescales: keV sterile neutrinos can indeed radiatively decay into a photon and a light active neutrino~\cite{Pal:1981rm}, with a lifetime proportional to the active-sterile mixing angle, that one can write as
\begin{equation}
	\theta_1^2 = \sum_\alpha \left|\left(U_\mathrm{PMNS}^* \sqrt{\hat{m}}\ R\right)_{\alpha 1}\right|^2 M_1^{-1}.
\end{equation}
Even if the sterile neutrino lifetime exceeds the age of the Universe, a fraction of DM can decay at present time, resulting in a monochromatic gamma-ray line with energy $M_1/2$ (i.e. X-rays). The non-observation of this signal results in an upper bound on $\theta_1^2$ as a function of the DM mass $M_1$~\cite{Boyarsky:2005us,Boyarsky:2006fg,Watson:2006qb,Yuksel:2007xh,Boyarsky:2007ge,Loewenstein:2008yi,RiemerSorensen:2009jp,Mirabal:2010an,Essig:2013goa,Horiuchi:2013noa,Riemer-Sorensen:2014yda,Tamura:2014mta,Ng:2015gfa,Riemer-Sorensen:2015kqa,Neronov:2016wdd,Perez:2016tcq,Dessert:2018qih,Boyarsky:2020hqb,Ng:2019gch,Roach:2019ctw,DeRomeri:2020wng,Foster:2021ngm}.

Finally, the DM momentum distribution affects the structure formation of the Universe: sterile neutrinos produced via the DW mechanism are classified as WDM, meaning they have a sizeable free streaming length that erases substructures at  galactic scales. A WDM dominated Universe is at odd with observation, while mixed scenarios where WDM constitutes a fraction $f_\mathrm{WDM}$ of the current DM relic density are potentially allowed, depending on the value of $f_\mathrm{WDM}$ and on the free-streaming length of the WDM component. In this study we adapt the 95\% C.L. results from~\cite{Baur:2017stq}, whose bounds can be translated for the case of sterile neutrinos produced via DW. Sterile neutrinos produced via freeze-in have a colder spectrum, compatible with the structure formation constraints for masses approximately above few keV~\cite{Bezrukov:2014nza,Abada:2014zra,Boyarsky:2018tvu} (see also~\cite{Shaposhnikov:2006xi,Merle:2015oja,Heeck:2017xbu}). In the present Letter we consider DM masses as light as 1 keV, keeping in mind that the validation of solutions with masses smaller than approximately 10 keV requires a dedicated study, that goes beyond the scope of the present work.
Finally, a lower bound on fermionic DM mass resulting from the phase space density of DM in dwarf spheroidal galaxies~\cite{Tremaine:1979we} excludes (disfavours) values below 400 eV (2 keV) (see~\cite{Boyarsky:2018tvu} and references therein).

\subsubsection{Laboratory}
We require the model in Eq.~(\ref{eq:seesaw_lag}) to correctly reproduce neutrino oscillation data~\cite{Esteban:2020cvm} at the 3$\sigma$ accuracy level. In addition, we impose constraints on the two heavy neutral leptons (HNL) from the non-observation of further beyond the SM signals: these include upper bounds on the active-sterile mixing from direct searches at accelerators~\cite{Das:2014jxa,Das:2015toa,Klinger:2014vdo,Sirunyan:2018mtv}, deviations from unitarity of the leptonic mixing matrix~\cite{Fernandez-Martinez:2016lgt} and neutrinoless double beta decay ($0\nu\beta\beta$)~\cite{Shirai:2017jyz}.

\subsubsection{Theory}
The CI parametrisation in Eq.~(\ref{eq:CI}) ensures that neutrino oscillation parameters are correctly reproduced. However, since the Yukawa couplings grow exponentially with the imaginary parts of the angles $\omega_{ij}$, these cannot be arbitrarily large, either because the theory becomes non-perturbative or because the Seesaw hypothesis $|vF|\ll |M|$ gets violated, invalidating the CI derivation itself. In our study, we impose an upper bound of $4\pi$ on each Yukawa coupling $|F_{\alpha I}|$, and require in addition that $\Gamma_I<M_I/2$ for each state~\cite{Chanowitz:1978mv,Durand:1989zs,Korner:1992an,Bernabeu:1993up,Fajfer:1998px,Ilakovac:1999md}. Moreover, for each realisation of the Dirac matrix~(\ref{eq:CI}) we explicitly diagonalise the full Lagrangian~(\ref{eq:seesaw_lag}), excluding solutions not reproducing neutrino oscillation data at the $3\sigma$ level.
Finally, we require the largest Yukawa coupling for each heavy neutrino to be larger than $\sqrt{2}\cdot 10^{-7}$~\cite{Akhmedov:1998qx}, so as to ensure that the state thermalises in the early Universe (a necessary condition for freeze-in production).

\begin{figure*}[htbp!]
\includegraphics[height=0.36\textwidth]{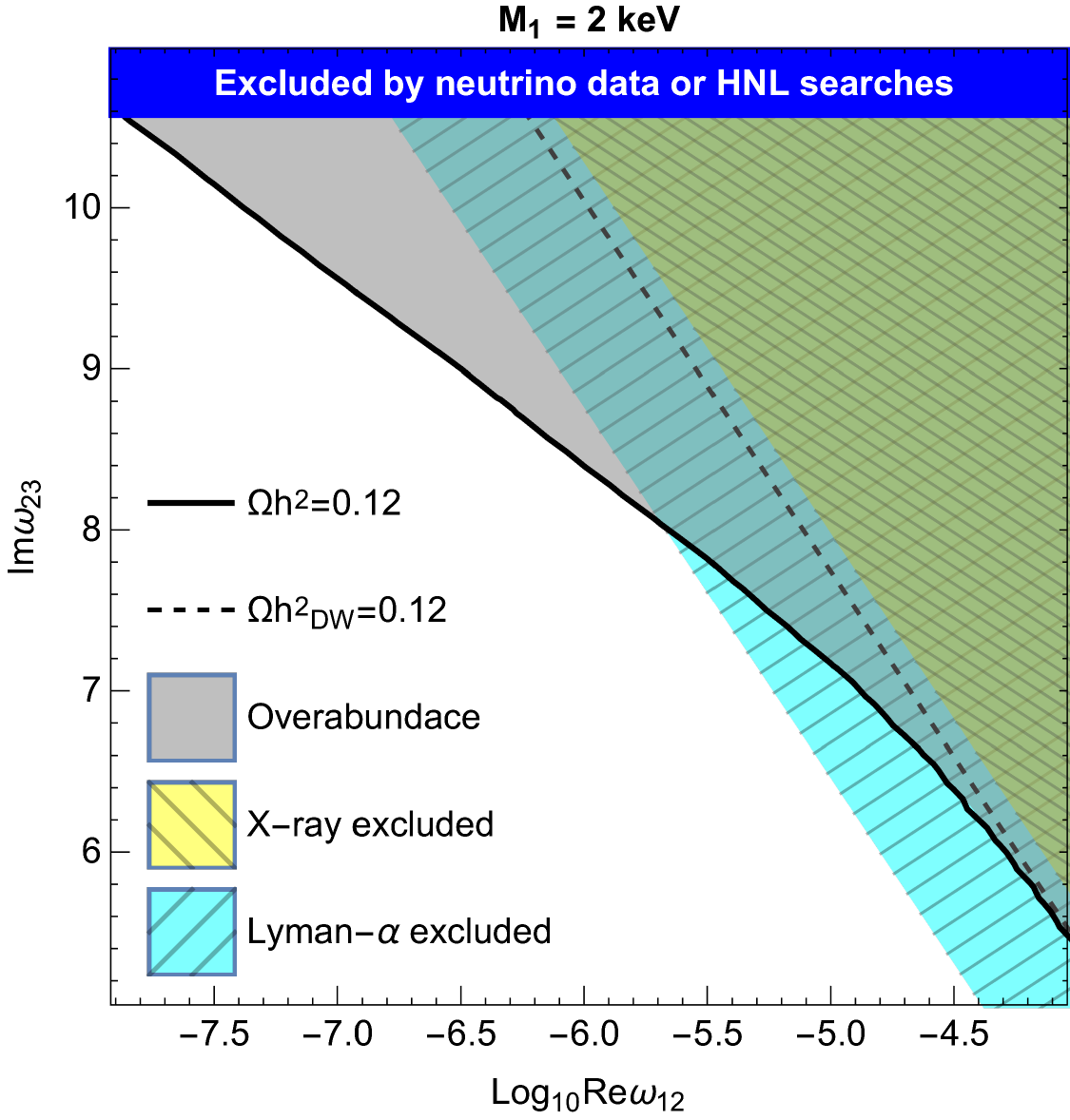}
\includegraphics[height=0.36\textwidth]{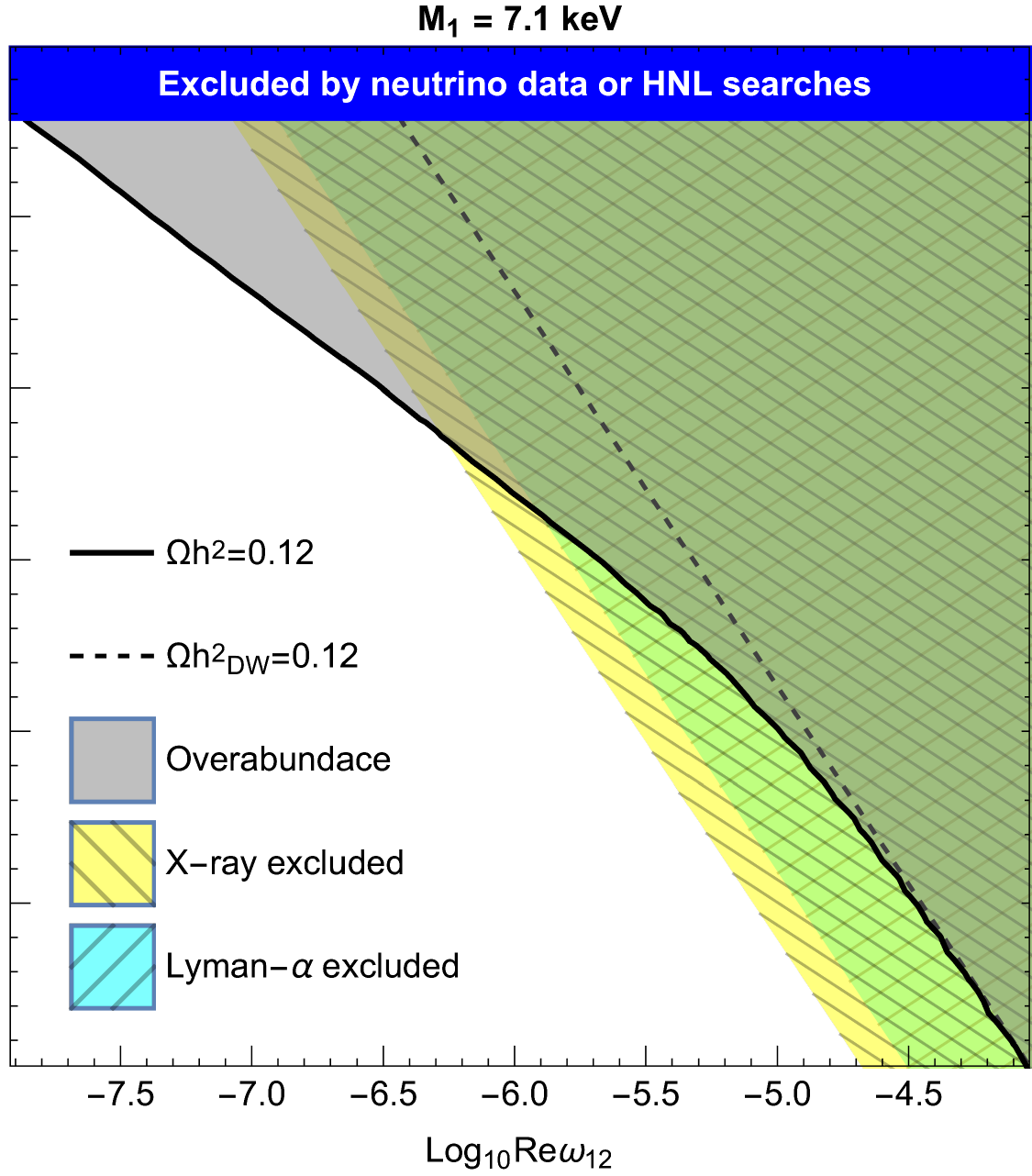}
\includegraphics[height=0.36\textwidth]{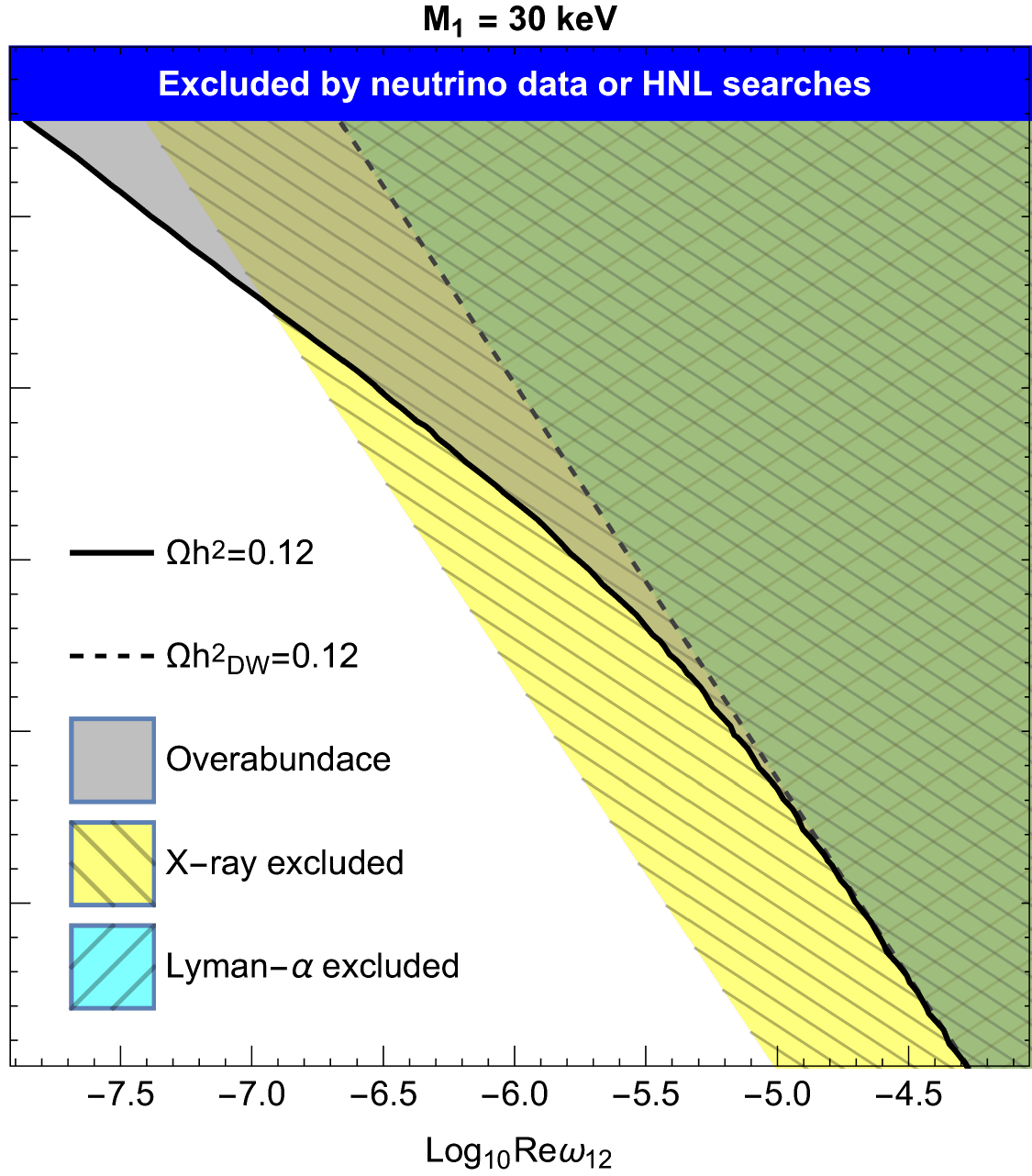}
\caption{\label{Fig:resultsNH} Parameter space of solutions for the model with the choice of parameters as described in the main text, for different DM masses. The black line (grey region) reproduces (exceeds) the observed relic density. Cyan and yellow regions are excluded by Lyman-$\alpha$ and X-ray observations, respectively, while in the blue region the model fails in reproducing neutrino oscillation data or is excluded by HNL searches. The black-dashed line shows the expected solution by considering the DW production mechanism only.}
\end{figure*}
\section{Methodology}\label{sec:methodology}
We chose to fix the neutrino oscillation parameters (mass differences, mixing angles and Dirac CP-violating phase) to their best-fit values in~\cite{Esteban:2020cvm}, and we set the CP-violating Majorana phases for the light neutrino states to zero, as their impact on the addressed dynamics is expected to be subdominant.
To enhance the predictivity of the model, we set to zero the lighter neutrino mass as well as the complex angle $\omega_{13}$, and assume that $\omega_{12}$ and $\omega_{23}$ are purely real and imaginary numbers, respectively. Together with fixing the heavy neutrino masses to 300 GeV, this leaves only 3 real free parameters: $M_1$, $\Re \omega_{12}$ and $\Im \omega_{23}$. This choice allows us to gain useful insights on the role played by the different parameters in the model.

\section{Results}\label{sec:results}
We report in Fig.~\ref{Fig:resultsNH} the parameter space of solutions for a Normal Ordering (NO) of light active neutrino masses, and three choices of DM mass: one where Lyman-$\alpha$ bounds are dominant, $M_1 = 2$ keV, one that reproduces the (not yet confirmed) monochromatic gamma-ray line at 3.55 keV~\cite{Bulbul:2014sua,Boyarsky:2014jta} ($M_1 = 7.1$ keV), and finally one where X-ray bounds are dominant, $M_1=30$ keV. In the plots, the black lines represent the choices of parameters that reproduce the observed DM relic abundance, $\Omega_\mathrm{DM} h^2 = 0.12$, while the grey region above is excluded by the overproduction of DM. The cyan and yellow regions are excluded by Lyman-$\alpha$ and X-ray observations, respectively, while in the blue region (on the top) the model does not reproduce neutrino oscillation data or is excluded by the HNL searches previously described: we find that the most constraining bounds in the present scenario are the ones imposed by the HNL contributions to charged lepton flavour violating processes, parametrised as deviations from unitarity of the leptonic mixing matrix (cf. e.g.~\cite{Fernandez-Martinez:2015hxa,Fernandez-Martinez:2016lgt} and references therein). The black-dashed line represents the expected solution if the freeze-in production is neglected, i.e. if all the relic abundance is due to the DW production: as it is well known, this solution is at odds with astrophysical observations. We also notice that the freeze-in production is less effective for decreasing values of $\Im \omega_{23}$.
The results for Inverted Ordering (IO) of light active neutrino masses are qualitatively similar, but this scenario corresponds to larger Yukawa couplings for fixed CI parameters, cf. Eq.~(\ref{eq:CI}). This implies that the upper bound on $\Im\omega_{23}$ from HNL constraints is stronger, $\Im\omega_{23}<9.87$ (while $\Im\omega_{23}<10.56$ for NO).

By increasing the value of the DM mass, X-ray constraints on $\theta^2_1$ become increasingly stronger (given that the sterile neutrino decay width is proportional to the fifth power of its mass): we find that in the considered scenario this closes the parameter space of solutions for masses $M_1$ above 60 keV (49 keV) for NO (IO).

\begin{figure*}[htb]
\includegraphics[height=0.21\textwidth]{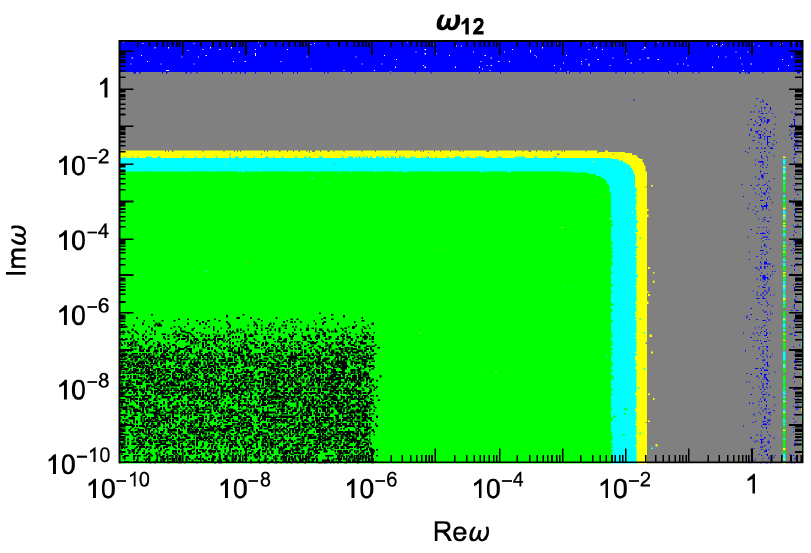}
\includegraphics[height=0.21\textwidth]{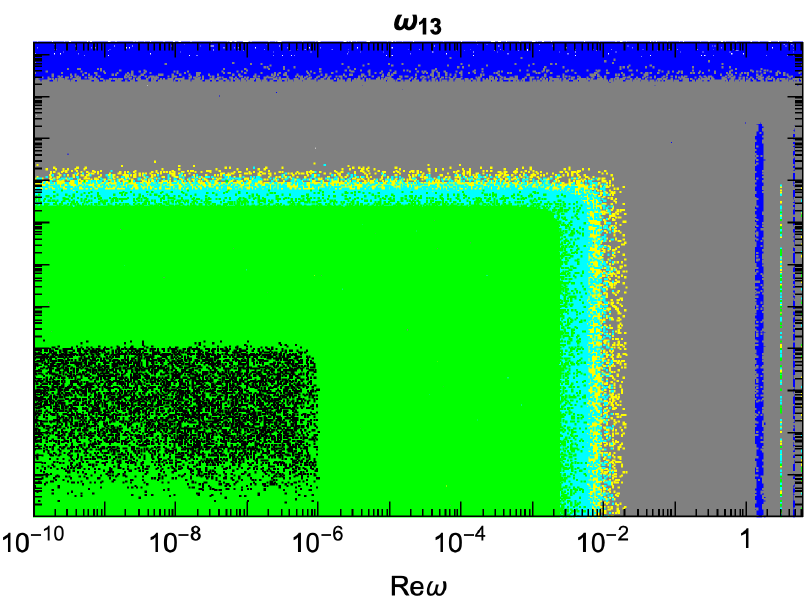}
\includegraphics[height=0.21\textwidth]{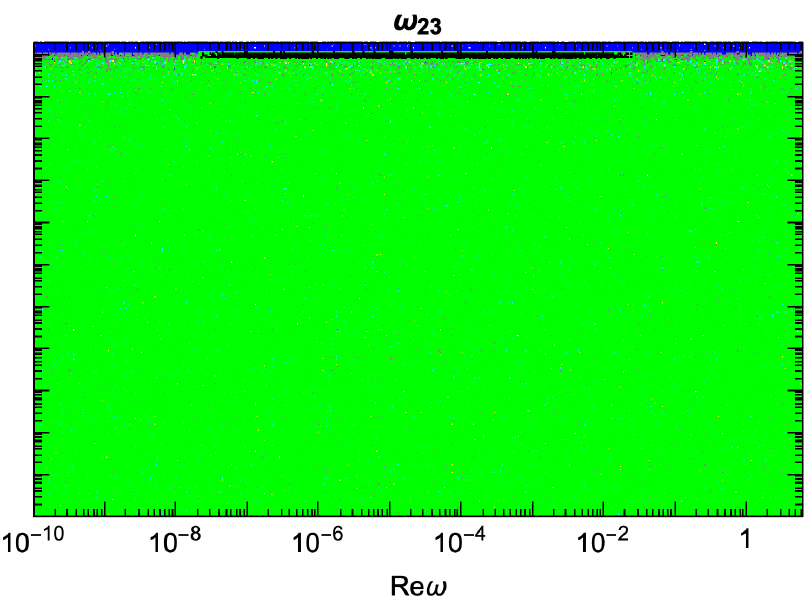}
\raisebox{0.8cm}{\includegraphics[width=0.1\textwidth]{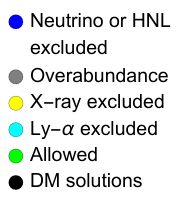}}
\caption{\label{fig:parameter_space} Allowed and excluded regions for the complex angles $\omega_{ij}$ in the CI parametrisation, with the other parameters chosen as described in the main text: green points are allowed by experimental constraints and theoretical considerations, while other colours denote exclusion by at least one of the criteria reported in the Figure. Finally, black points reproduce the observed DM relic density at the $3\sigma$ precision.}
\end{figure*}
\subsection{Phenomenologically emergent approximate lepton number symmetry}
The parameter space of solutions discussed in the previous section may seem strongly fine tuned, featuring a large hierarchy of values between the DM mass and the heavier HNLs, as well as between  $\Im \omega_{23}$ and the other parameters of the $R$ matrix. However, this is exactly the limit in which global lepton number conservation is recovered in the CI parametrisation: in the limit $M_1 \rightarrow 0$, $M_2 \rightarrow M_3$, $\omega_{12}, \omega_{13} \rightarrow 0$ and $\omega_{23} \rightarrow \pm i \infty$, the new fields arrange in a decoupled massless state $N_1$ and a Dirac fermion given by a linear combination of the fields $N_2$ and $N_3$, thus resulting in a theory with lepton number conserving interactions only. In the exactly symmetric limit light active neutrinos are massless, $\hat{m}=0$, thus the symmetry must be broken at some level. This also provides a possible explanation for the lightness of active neutrinos when compared to the other SM fermions~\cite{tHooft:1979rat,Gluza:2002vs,Shaposhnikov:2006nn,Kersten:2007vk,Abada:2007ux,Moffat:2017feq}.
Concerning the specific values of $M_i$, the symmetry only imposes relations among them, while the absolute scales $M_1$ and $M_{2,3}$ must derive from the (unknown) mechanism at the origin of these Majorana mass terms. Without assuming a specific mechanism, the restrictions on these parameters are mostly phenomenological in nature. In particular, to be a viable DM candidate $M_1$ must lie around the keV scale (smaller values are excluded by structure formation, larger ones give a too short DM lifetime), while the scale of $M_{2,3}$ is determined by the assumed DM production mechanism (much smaller values imply too small Yukawa couplings to produce a sizeable DM population, while for much larger values the parent particles decay at temperatures above the electroweak phase transition, when the active-sterile mixings are vanishing).

We report in Fig.~\ref{fig:parameter_space} the results of a general scan on the parameter space of the model, performed for NO, fixing the DM mass $M_1=5$ keV and randomly varying the parameters $\Re \omega_{ij}$ and $\Im \omega_{ij}$  in the range $[10^{-10},2\pi]$ and $[10^{-10},20]$ respectively, drawing from a log-flat distribution. We also allow the physical Majorana phase in the active neutrino sector to vary in the $\left[0,2\pi\right]$ range.  We conclude that astrophysical constraints set an upper bound $\left|\omega_{1i}\right| \lesssim 6\cdot 10^{-3}$ ($i=2,3$) (clearly, values of $\Re\omega_{1i}$ above $(2\pi - 10^{-3})$ are as well allowed), while no bound is identified for $\Re \omega_{23}$ and the upper bound on $\Im \omega_{23}$ ($\Im \omega_{23} \lesssim 11$) is solely determined by the requirement of not overproducing the DM relic density. We thus conclude that experimental constraints force the parameters $\omega_{12}$ and $\omega_{13}$ to be much smaller than unity. In addition, the requirement of reproducing the observed DM relic abundance via freeze-in decay requires HNLs heavier than the electroweak scale, thus introducing a hierarchy of masses, as well as $\Im \omega_{23} \approx 10$, recovering the approximate symmetric limit. Concerning the mass degeneracy between $M_{2,3}$, this suppresses the $0\nu\beta\beta$ decay rate\footnote{We thank Jordy de Vries for having encouraged us to explicitly verify that very large mass splittings in this region of the parameter space can result in $0\nu\beta\beta$ rates exceeding current experimental bounds.} which could be sizeable for large values of $\Im\omega_{23}$, although mild mass splittings of $\mathcal{O}(10)$ GeV are nevertheless allowed; relaxing this condition relaxes the symmetry, thus resulting in a less justified framework from the model building point of view, while opening at the same time to the possibility of large loop corrections to the active neutrino masses~\cite{Lopez-Pavon:2015cga}.
The approximate lepton number symmetry is known to play a role in justifying the structure of the $\nu$MSM~\cite{Shaposhnikov:2006nn}, as well as in other low-scale leptogenesis models~\cite{Abada:2015rta,Abada:2017ieq,Abada:2018oly}.

\section{Conclusion}
We have shown that, in the minimal SM extended with 3 RHN, the freeze-in production of a sterile neutrino DM can account for the observed relic abundance, while complying with existing constraints. This production channel has been, to the best of our knowledge, neglected in previous studies, and effectively opens a new parameter space of solutions for the sterile neutrino DM hypothesis.
This  space of solutions is characterised by a hierarchy between different parameters of the model (sterile neutrino masses as well as complex angles in the CI parametrisation) that points towards an approximate lepton number symmetry, which could be embedded in a larger $B-L$ framework.
When presenting solutions, we have worked under a simplifying assumption in order to reduce the number of free parameters and gain a deeper insight on the model; the actual parameter space of solutions is thus larger than the one presented here, and we postpone a dedicated statistically comprehensive study to a future work.

\section*{Acknowledgments}
I am deeply grateful to Asmaa Abada, Giorgio Arcadi, Gioacchino Piazza and Salvador Rosauro-Alcaraz for insightful discussions on the impact of thermal effects on the production mechanisms.
I thank Marco Drewes for useful discussions on the connection between the CI parametrisation and global lepton number symmetry, as well as Asmaa Abada and Felix Kahlhoefer for useful comments on the manuscript. This work is funded by the European Union under the Horizon Europe's Marie Sklodowska-Curie project 101068791 - NuBridge, and has been supported by the Alexander von Humboldt Foundation.

\appendix

\section{Explicit numerical example}
We report hereafter the numerical values for an example point that reproduces the correct DM abundance via Eq.s~{(\ref{eq:DW}, \ref{eq:FI_relic})}.
	
The input values in the CI parametrisation are

\begin{eqnarray}
	m_\nu^{\textrm{\tiny{lightest}}} &=& 0, \\
	M_1 &=& 5\ \textrm{keV},\\
	M_2 &=& 300\ \textrm{GeV},\\
	M_3 &=& 300\ \textrm{GeV},\\
	\omega_{12} &=& 5.184\times 10^{-10}+ i\ 3.867\times 10^{-8} ,\\
	\omega_{13} &=& 9.678\times 10^{-7}+ i\ 6.241\times 10^{-8},\\
	\omega_{23} &=& 2.439\times
   10^{-5}+ i\ 8.398, \\
	\alpha_1 &=& 3.763, \\
	\alpha_2 &=& 2.021,
\end{eqnarray}
while the light neutrino mass differences and PMNS parameters are fixed to the best fit values from~\cite{Esteban:2020cvm}.

With this input the following mass matrix results
\begin{equation}
 \mathcal{M} = \left(
 	\begin{array}{cc}
 		0 & m_D \\
 		m_D^T & M_M 
 	\end{array}
 \right)\ \textrm{GeV},
\end{equation}
with
\begin{equation}
	m_D = \left(
\begin{array}{ccc}
 \mathbf{m_D^1} & \mathbf{m_D^2} & \mathbf{m_D^3}
\end{array}
\right),
\end{equation}
and the individual vectors are given by
\begin{eqnarray}
	\mathbf{m_D^1} &=& \left(
\begin{array}{c}
 -1.233\times 10^{-12}+ i\ 3.558\times 10^{-12} \\
 2.401\times 10^{-11}+ i\ 2.399\times 10^{-11} \\
 1.673\times 10^{-11}+ i\ 5.020\times 10^{-12} \\
\end{array}
\right), \\
\mathbf{m_D^2} &=& \left(
\begin{array}{c}
 -2.786\times 10^{-2}- i\ 7.779\times 10^{-3} \\
 -1.726\times 10^{-1}+ i\ 1.954\times 10^{-1}\\
 -3.060\times 10^{-2}+ i\ 1.307\times 10^{-1} \\
\end{array}
\right), \\
\mathbf{m_D^3} &=& \left(
\begin{array}{c}
 7.779\times 10^{-3}- i \ 2.786\times 10^{-2} \\
 -1.954\times 10^{-1}- i\ 1.726\times 10^{-1} \\
 -1.307\times 10^{-1}- i\ 3.060\times 10^{-2} \\
\end{array}
\right),
 \\
M_M &=& \left(
\begin{array}{ccc}
 5.\times 10^{-6} & 0. & 0. \\
 0. & 3.\times 10^2 & 0. \\
 0. & 0. & 3.\times 10^2 \\
\end{array}
\right).
\end{eqnarray}

Upon diagonalisation of $\mathcal{M}$, the following values for the mass eigenstates are obtained
\begin{equation}
\left(
\begin{array}{c}
 0 \\
 8.6139397443927\times 10^{-3} \\
 5.0169629398723\times 10^{-2} \\
 5.0000000002942\times 10^3 \\
 3.0000028935617\times 10^{11} \\
 3.0000028935621\times 10^{11} \\
\end{array}
\right) \ \textrm{eV},
\end{equation}
with the PMNS matrix (i.e. the $3\times3$ upper-left block of the full mixing matrix $\mathcal{U}$)
\begin{equation}
U_\textrm{PMNS} = \left(
\begin{array}{ccc}
 0.83 & 0.17-0.52 i & -0.11-0.1 i \\
 -0.27+0.03 i & 0.20-0.57 i & 0.40+0.63 i \\
 0.49+0.02 i & -0.16+0.56 i & 0.34+0.55 i \\
\end{array}
\right),
\end{equation}
and the following values for the active-sterile mixings (i.e. the elements in the upper-right block of $\mathcal{U}$)
\begin{eqnarray}
	\mathbf{\mathcal{U}_{\alpha 4}} &=& \left(
\begin{array}{c}
 -2.466\times 10^{-7}- i\ 7.115\times 10^{-7} \\
 4.801\times 10^{-6}- i\ 4.797\times 10^{-6} \\
 3.346\times 10^{-6}- i\ 1.004\times 10^{-6} \\
\end{array}
\right), \\
\mathbf{\mathcal{U}_{\alpha 5}} &=& \left(
\begin{array}{c}
 -9.286\times 10^{-5}+ i\ 2.593\times 10^{-5}\\
 -5.753\times 10^{-4}- i\ 6.512\times 10^{-4} \\
 -1.020\times 10^{-4}- i\ 4.355\times 10^{-4}\\
\end{array}
\right), \\
\mathbf{\mathcal{U}_{\alpha 6}}	&=& \left(
\begin{array}{c}
 2.593\times 10^{-5}+ i\ 9.286\times 10^{-5} \\
 -6.512\times 10^{-4}+ i\ 5.753\times 10^{-4} \\
 -4.355\times 10^{-4}+ i\ 1.020\times 10^{-4} \\
\end{array}
\right).
\end{eqnarray}

Following Eq.s~{(\ref{eq:DW}, \ref{eq:FI_relic})}, this point gives a total DM abundance of
\begin{equation}
	\Omega h^2 = 0.1173,
\end{equation}
composed by $\Omega^\textrm{\tiny FI} h^2 = 0.1081$ from freeze-in decay, and $\Omega^\textrm{\tiny DW} h^2 = 0.0092$ from Dodelson-Widrow production.

One can notice in above equations the approximate lepton-number conserving structure, with $M_1 \ll M_{2,3}$, $M_2 \sim M_3$, ${m_D^1} \ll {m_D^{2,3}}$ and $m_D^3 \simeq i \ m_D^2$.
\\

We stress that when numerically diagonalising such kind of solutions it is important to use a sufficiently large precision, since in a symmetry-protected scenario the sub-eV neutrino masses can result from cancellations between terms at much larger energy scales (such cancellations are indeed imposed by the symmetry itself).

\bibliographystyle{elsarticle-num} 
\bibliography{SeesawFIDM_v2.bib}

\begin{thebibliography}{10}
\expandafter\ifx\csname url\endcsname\relax
  \def\url#1{\texttt{#1}}\fi
\expandafter\ifx\csname urlprefix\endcsname\relax\def\urlprefix{URL }\fi
\expandafter\ifx\csname href\endcsname\relax
  \def\href#1#2{#2} \def\path#1{#1}\fi

\bibitem{Minkowski:1977sc}
P.~Minkowski, {$\mu \to e\gamma$ at a Rate of One Out of $10^{9}$ Muon Decays?}, Phys. Lett. B 67 (1977) 421--428.
\newblock \href {http://dx.doi.org/10.1016/0370-2693(77)90435-X} {\path{doi:10.1016/0370-2693(77)90435-X}}.

\bibitem{GellMann:1980vs}
M.~Gell-Mann, P.~Ramond, R.~Slansky, {Complex Spinors and Unified Theories}, Conf. Proc. C 790927 (1979) 315--321.
\newblock \href {http://arxiv.org/abs/1306.4669} {\path{arXiv:1306.4669}}.

\bibitem{Mohapatra:1979ia}
R.~N. Mohapatra, G.~Senjanovic, {Neutrino Mass and Spontaneous Parity Nonconservation}, Phys. Rev. Lett. 44 (1980) 912.
\newblock \href {http://dx.doi.org/10.1103/PhysRevLett.44.912} {\path{doi:10.1103/PhysRevLett.44.912}}.

\bibitem{Yanagida:1980xy}
T.~Yanagida, {Horizontal Symmetry and Masses of Neutrinos}, Prog. Theor. Phys. 64 (1980) 1103.
\newblock \href {http://dx.doi.org/10.1143/PTP.64.1103} {\path{doi:10.1143/PTP.64.1103}}.

\bibitem{Schechter:1980gr}
J.~Schechter, J.~W.~F. Valle, {Neutrino Masses in SU(2) x U(1) Theories}, Phys. Rev. D 22 (1980) 2227.
\newblock \href {http://dx.doi.org/10.1103/PhysRevD.22.2227} {\path{doi:10.1103/PhysRevD.22.2227}}.

\bibitem{Schechter:1981cv}
J.~Schechter, J.~W.~F. Valle, {Neutrino Decay and Spontaneous Violation of Lepton Number}, Phys. Rev. D 25 (1982) 774.
\newblock \href {http://dx.doi.org/10.1103/PhysRevD.25.774} {\path{doi:10.1103/PhysRevD.25.774}}.

\bibitem{Mohapatra:1998rq}
R.~N. Mohapatra, P.~B. Pal, {Massive neutrinos in physics and astrophysics. Second edition}, Vol.~60, 1998.

\bibitem{Boyarsky:2018tvu}
A.~Boyarsky, M.~Drewes, T.~Lasserre, S.~Mertens, O.~Ruchayskiy, {Sterile neutrino Dark Matter}, Prog. Part. Nucl. Phys. 104 (2019) 1--45.
\newblock \href {http://arxiv.org/abs/1807.07938} {\path{arXiv:1807.07938}}, \href {http://dx.doi.org/10.1016/j.ppnp.2018.07.004} {\path{doi:10.1016/j.ppnp.2018.07.004}}.

\bibitem{Aghanim:2018eyx}
N.~Aghanim, et~al., {Planck 2018 results. VI. Cosmological parameters}, Astron. Astrophys. 641 (2020) A6, [Erratum: Astron.Astrophys. 652, C4 (2021)].
\newblock \href {http://arxiv.org/abs/1807.06209} {\path{arXiv:1807.06209}}, \href {http://dx.doi.org/10.1051/0004-6361/201833910} {\path{doi:10.1051/0004-6361/201833910}}.

\bibitem{Pal:1981rm}
P.~B. Pal, L.~Wolfenstein, {Radiative Decays of Massive Neutrinos}, Phys. Rev. D 25 (1982) 766.
\newblock \href {http://dx.doi.org/10.1103/PhysRevD.25.766} {\path{doi:10.1103/PhysRevD.25.766}}.

\bibitem{Tremaine:1979we}
S.~Tremaine, J.~E. Gunn, {Dynamical Role of Light Neutral Leptons in Cosmology}, Phys. Rev. Lett. 42 (1979) 407--410.
\newblock \href {http://dx.doi.org/10.1103/PhysRevLett.42.407} {\path{doi:10.1103/PhysRevLett.42.407}}.

\bibitem{Boyarsky:2008ju}
A.~Boyarsky, O.~Ruchayskiy, D.~Iakubovskyi, {A Lower bound on the mass of Dark Matter particles}, JCAP 03 (2009) 005.
\newblock \href {http://arxiv.org/abs/0808.3902} {\path{arXiv:0808.3902}}, \href {http://dx.doi.org/10.1088/1475-7516/2009/03/005} {\path{doi:10.1088/1475-7516/2009/03/005}}.

\bibitem{Hui:1996fh}
L.~Hui, N.~Y. Gnedin, Y.~Zhang, {The Statistics of density peaks and the column density distribution of the Lyman-alpha forest}, Astrophys. J. 486 (1997) 599.
\newblock \href {http://arxiv.org/abs/astro-ph/9608157} {\path{arXiv:astro-ph/9608157}}, \href {http://dx.doi.org/10.1086/304539} {\path{doi:10.1086/304539}}.

\bibitem{Gnedin:2001wg}
N.~Y. Gnedin, A.~J.~S. Hamilton, {Matter power spectrum from the Lyman-alpha forest: Myth or reality?}, Mon. Not. Roy. Astron. Soc. 334 (2002) 107--116.
\newblock \href {http://arxiv.org/abs/astro-ph/0111194} {\path{arXiv:astro-ph/0111194}}, \href {http://dx.doi.org/10.1046/j.1365-8711.2002.05490.x} {\path{doi:10.1046/j.1365-8711.2002.05490.x}}.

\bibitem{Weinberg:2003eg}
D.~H. Weinberg, R.~Dave, N.~Katz, J.~A. Kollmeier, {The Lyman - alpha forest as a cosmological tool}, AIP Conf. Proc. 666~(1) (2003) 157--169.
\newblock \href {http://arxiv.org/abs/astro-ph/0301186} {\path{arXiv:astro-ph/0301186}}, \href {http://dx.doi.org/10.1063/1.1581786} {\path{doi:10.1063/1.1581786}}.

\bibitem{Boyarsky:2008xj}
A.~Boyarsky, J.~Lesgourgues, O.~Ruchayskiy, M.~Viel, {Lyman-alpha constraints on warm and on warm-plus-cold dark matter models}, JCAP 05 (2009) 012.
\newblock \href {http://arxiv.org/abs/0812.0010} {\path{arXiv:0812.0010}}, \href {http://dx.doi.org/10.1088/1475-7516/2009/05/012} {\path{doi:10.1088/1475-7516/2009/05/012}}.

\bibitem{Dodelson:1993je}
S.~Dodelson, L.~M. Widrow, {Sterile-neutrinos as dark matter}, Phys. Rev. Lett. 72 (1994) 17--20.
\newblock \href {http://arxiv.org/abs/hep-ph/9303287} {\path{arXiv:hep-ph/9303287}}, \href {http://dx.doi.org/10.1103/PhysRevLett.72.17} {\path{doi:10.1103/PhysRevLett.72.17}}.

\bibitem{Baur:2017stq}
J.~Baur, N.~Palanque-Delabrouille, C.~Yeche, A.~Boyarsky, O.~Ruchayskiy, E.~Armengaud, J.~Lesgourgues, {Constraints from Ly-$\alpha$ forests on non-thermal dark matter including resonantly-produced sterile neutrinos}, JCAP 12 (2017) 013.
\newblock \href {http://arxiv.org/abs/1706.03118} {\path{arXiv:1706.03118}}, \href {http://dx.doi.org/10.1088/1475-7516/2017/12/013} {\path{doi:10.1088/1475-7516/2017/12/013}}.

\bibitem{Boyarsky:2005us}
A.~Boyarsky, A.~Neronov, O.~Ruchayskiy, M.~Shaposhnikov, {Constraints on sterile neutrino as a dark matter candidate from the diffuse x-ray background}, Mon. Not. Roy. Astron. Soc. 370 (2006) 213--218.
\newblock \href {http://arxiv.org/abs/astro-ph/0512509} {\path{arXiv:astro-ph/0512509}}, \href {http://dx.doi.org/10.1111/j.1365-2966.2006.10458.x} {\path{doi:10.1111/j.1365-2966.2006.10458.x}}.

\bibitem{Boyarsky:2006fg}
A.~Boyarsky, A.~Neronov, O.~Ruchayskiy, M.~Shaposhnikov, I.~Tkachev, {Where to find a dark matter sterile neutrino?}, Phys. Rev. Lett. 97 (2006) 261302.
\newblock \href {http://arxiv.org/abs/astro-ph/0603660} {\path{arXiv:astro-ph/0603660}}, \href {http://dx.doi.org/10.1103/PhysRevLett.97.261302} {\path{doi:10.1103/PhysRevLett.97.261302}}.

\bibitem{Watson:2006qb}
C.~R. Watson, J.~F. Beacom, H.~Yuksel, T.~P. Walker, {Direct X-ray Constraints on Sterile Neutrino Warm Dark Matter}, Phys. Rev. D 74 (2006) 033009.
\newblock \href {http://arxiv.org/abs/astro-ph/0605424} {\path{arXiv:astro-ph/0605424}}, \href {http://dx.doi.org/10.1103/PhysRevD.74.033009} {\path{doi:10.1103/PhysRevD.74.033009}}.

\bibitem{Yuksel:2007xh}
H.~Yuksel, J.~F. Beacom, C.~R. Watson, {Strong Upper Limits on Sterile Neutrino Warm Dark Matter}, Phys. Rev. Lett. 101 (2008) 121301.
\newblock \href {http://arxiv.org/abs/0706.4084} {\path{arXiv:0706.4084}}, \href {http://dx.doi.org/10.1103/PhysRevLett.101.121301} {\path{doi:10.1103/PhysRevLett.101.121301}}.

\bibitem{Boyarsky:2007ge}
A.~Boyarsky, D.~Malyshev, A.~Neronov, O.~Ruchayskiy, {Constraining DM properties with SPI}, Mon. Not. Roy. Astron. Soc. 387 (2008) 1345.
\newblock \href {http://arxiv.org/abs/0710.4922} {\path{arXiv:0710.4922}}, \href {http://dx.doi.org/10.1111/j.1365-2966.2008.13003.x} {\path{doi:10.1111/j.1365-2966.2008.13003.x}}.

\bibitem{Loewenstein:2008yi}
M.~Loewenstein, A.~Kusenko, P.~L. Biermann, {New Limits on Sterile Neutrinos from Suzaku Observations of the Ursa Minor Dwarf Spheroidal Galaxy}, Astrophys. J. 700 (2009) 426--435.
\newblock \href {http://arxiv.org/abs/0812.2710} {\path{arXiv:0812.2710}}, \href {http://dx.doi.org/10.1088/0004-637X/700/1/426} {\path{doi:10.1088/0004-637X/700/1/426}}.

\bibitem{RiemerSorensen:2009jp}
S.~Riemer-Sorensen, S.~H. Hansen, {Decaying dark matter in Draco}, Astron. Astrophys. 500 (2009) L37--L40.
\newblock \href {http://arxiv.org/abs/0901.2569} {\path{arXiv:0901.2569}}, \href {http://dx.doi.org/10.1051/0004-6361/200912430} {\path{doi:10.1051/0004-6361/200912430}}.

\bibitem{Mirabal:2010an}
N.~Mirabal, {Swift observation of Segue 1: constraints on sterile neutrino parameters in the darkest galaxy}, Mon. Not. Roy. Astron. Soc. 409 (2010) 128.
\newblock \href {http://arxiv.org/abs/1010.4706} {\path{arXiv:1010.4706}}, \href {http://dx.doi.org/10.1111/j.1745-3933.2010.00963.x} {\path{doi:10.1111/j.1745-3933.2010.00963.x}}.

\bibitem{Essig:2013goa}
R.~Essig, E.~Kuflik, S.~D. McDermott, T.~Volansky, K.~M. Zurek, {Constraining Light Dark Matter with Diffuse X-Ray and Gamma-Ray Observations}, JHEP 11 (2013) 193.
\newblock \href {http://arxiv.org/abs/1309.4091} {\path{arXiv:1309.4091}}, \href {http://dx.doi.org/10.1007/JHEP11(2013)193} {\path{doi:10.1007/JHEP11(2013)193}}.

\bibitem{Horiuchi:2013noa}
S.~Horiuchi, P.~J. Humphrey, J.~Onorbe, K.~N. Abazajian, M.~Kaplinghat, S.~Garrison-Kimmel, {Sterile neutrino dark matter bounds from galaxies of the Local Group}, Phys. Rev. D 89~(2) (2014) 025017.
\newblock \href {http://arxiv.org/abs/1311.0282} {\path{arXiv:1311.0282}}, \href {http://dx.doi.org/10.1103/PhysRevD.89.025017} {\path{doi:10.1103/PhysRevD.89.025017}}.

\bibitem{Riemer-Sorensen:2014yda}
S.~Riemer-S\o{}rensen, {Constraints on the presence of a 3.5 keV dark matter emission line from Chandra observations of the Galactic centre}, Astron. Astrophys. 590 (2016) A71.
\newblock \href {http://arxiv.org/abs/1405.7943} {\path{arXiv:1405.7943}}, \href {http://dx.doi.org/10.1051/0004-6361/201527278} {\path{doi:10.1051/0004-6361/201527278}}.

\bibitem{Tamura:2014mta}
T.~Tamura, R.~Iizuka, Y.~Maeda, K.~Mitsuda, N.~Y. Yamasaki, {An X-ray Spectroscopic Search for Dark Matter in the Perseus Cluster with Suzaku}, Publ. Astron. Soc. Jap. 67 (2015) 23.
\newblock \href {http://arxiv.org/abs/1412.1869} {\path{arXiv:1412.1869}}, \href {http://dx.doi.org/10.1093/pasj/psu156} {\path{doi:10.1093/pasj/psu156}}.

\bibitem{Ng:2015gfa}
K.~C.~Y. Ng, S.~Horiuchi, J.~M. Gaskins, M.~Smith, R.~Preece, {Improved Limits on Sterile Neutrino Dark Matter using Full-Sky Fermi Gamma-Ray Burst Monitor Data}, Phys. Rev. D 92~(4) (2015) 043503.
\newblock \href {http://arxiv.org/abs/1504.04027} {\path{arXiv:1504.04027}}, \href {http://dx.doi.org/10.1103/PhysRevD.92.043503} {\path{doi:10.1103/PhysRevD.92.043503}}.

\bibitem{Riemer-Sorensen:2015kqa}
S.~Riemer-S\o{}rensen, et~al., {Dark matter line emission constraints from NuSTAR observations of the Bullet Cluster}, Astrophys. J. 810~(1) (2015) 48.
\newblock \href {http://arxiv.org/abs/1507.01378} {\path{arXiv:1507.01378}}, \href {http://dx.doi.org/10.1088/0004-637X/810/1/48} {\path{doi:10.1088/0004-637X/810/1/48}}.

\bibitem{Neronov:2016wdd}
A.~Neronov, D.~Malyshev, D.~Eckert, {Decaying dark matter search with NuSTAR deep sky observations}, Phys. Rev. D 94~(12) (2016) 123504.
\newblock \href {http://arxiv.org/abs/1607.07328} {\path{arXiv:1607.07328}}, \href {http://dx.doi.org/10.1103/PhysRevD.94.123504} {\path{doi:10.1103/PhysRevD.94.123504}}.

\bibitem{Perez:2016tcq}
K.~Perez, K.~C.~Y. Ng, J.~F. Beacom, C.~Hersh, S.~Horiuchi, R.~Krivonos, {Almost closing the \ensuremath{\nu}MSM sterile neutrino dark matter window with NuSTAR}, Phys. Rev. D 95~(12) (2017) 123002.
\newblock \href {http://arxiv.org/abs/1609.00667} {\path{arXiv:1609.00667}}, \href {http://dx.doi.org/10.1103/PhysRevD.95.123002} {\path{doi:10.1103/PhysRevD.95.123002}}.

\bibitem{Dessert:2018qih}
C.~Dessert, N.~L. Rodd, B.~R. Safdi, {The dark matter interpretation of the 3.5-keV line is inconsistent with blank-sky observations}, Science 367~(6485) (2020) 1465--1467.
\newblock \href {http://arxiv.org/abs/1812.06976} {\path{arXiv:1812.06976}}, \href {http://dx.doi.org/10.1126/science.aaw3772} {\path{doi:10.1126/science.aaw3772}}.

\bibitem{Boyarsky:2020hqb}
A.~Boyarsky, D.~Malyshev, O.~Ruchayskiy, D.~Savchenko, {Technical comment on the paper of Dessert et al. ''The dark matter interpretation of the 3.5 keV line is inconsistent with blank-sky observations''}\href {http://arxiv.org/abs/2004.06601} {\path{arXiv:2004.06601}}.

\bibitem{Ng:2019gch}
K.~C.~Y. Ng, B.~M. Roach, K.~Perez, J.~F. Beacom, S.~Horiuchi, R.~Krivonos, D.~R. Wik, {New Constraints on Sterile Neutrino Dark Matter from $NuSTAR$ M31 Observations}, Phys. Rev. D 99 (2019) 083005.
\newblock \href {http://arxiv.org/abs/1901.01262} {\path{arXiv:1901.01262}}, \href {http://dx.doi.org/10.1103/PhysRevD.99.083005} {\path{doi:10.1103/PhysRevD.99.083005}}.

\bibitem{Roach:2019ctw}
B.~M. Roach, K.~C.~Y. Ng, K.~Perez, J.~F. Beacom, S.~Horiuchi, R.~Krivonos, D.~R. Wik, {NuSTAR Tests of Sterile-Neutrino Dark Matter: New Galactic Bulge Observations and Combined Impact}, Phys. Rev. D 101~(10) (2020) 103011.
\newblock \href {http://arxiv.org/abs/1908.09037} {\path{arXiv:1908.09037}}, \href {http://dx.doi.org/10.1103/PhysRevD.101.103011} {\path{doi:10.1103/PhysRevD.101.103011}}.

\bibitem{DeRomeri:2020wng}
V.~De~Romeri, D.~Karamitros, O.~Lebedev, T.~Toma, {Neutrino dark matter and the Higgs portal: improved freeze-in analysis}, JHEP 10 (2020) 137.
\newblock \href {http://arxiv.org/abs/2003.12606} {\path{arXiv:2003.12606}}, \href {http://dx.doi.org/10.1007/JHEP10(2020)137} {\path{doi:10.1007/JHEP10(2020)137}}.

\bibitem{Foster:2021ngm}
J.~W. Foster, M.~Kongsore, C.~Dessert, Y.~Park, N.~L. Rodd, K.~Cranmer, B.~R. Safdi, {Deep Search for Decaying Dark Matter with XMM-Newton Blank-Sky Observations}, Phys. Rev. Lett. 127~(5) (2021) 051101.
\newblock \href {http://arxiv.org/abs/2102.02207} {\path{arXiv:2102.02207}}, \href {http://dx.doi.org/10.1103/PhysRevLett.127.051101} {\path{doi:10.1103/PhysRevLett.127.051101}}.

\bibitem{Shi:1998km}
X.-D. Shi, G.~M. Fuller, {A New dark matter candidate: Nonthermal sterile neutrinos}, Phys. Rev. Lett. 82 (1999) 2832--2835.
\newblock \href {http://arxiv.org/abs/astro-ph/9810076} {\path{arXiv:astro-ph/9810076}}, \href {http://dx.doi.org/10.1103/PhysRevLett.82.2832} {\path{doi:10.1103/PhysRevLett.82.2832}}.

\bibitem{Asaka:2005an}
T.~Asaka, S.~Blanchet, M.~Shaposhnikov, {The nuMSM, dark matter and neutrino masses}, Phys. Lett. B 631 (2005) 151--156.
\newblock \href {http://arxiv.org/abs/hep-ph/0503065} {\path{arXiv:hep-ph/0503065}}, \href {http://dx.doi.org/10.1016/j.physletb.2005.09.070} {\path{doi:10.1016/j.physletb.2005.09.070}}.

\bibitem{Asaka:2005pn}
T.~Asaka, M.~Shaposhnikov, {The $\nu$MSM, dark matter and baryon asymmetry of the universe}, Phys. Lett. B 620 (2005) 17--26.
\newblock \href {http://arxiv.org/abs/hep-ph/0505013} {\path{arXiv:hep-ph/0505013}}, \href {http://dx.doi.org/10.1016/j.physletb.2005.06.020} {\path{doi:10.1016/j.physletb.2005.06.020}}.

\bibitem{Canetti:2012kh}
L.~Canetti, M.~Drewes, T.~Frossard, M.~Shaposhnikov, {Dark Matter, Baryogenesis and Neutrino Oscillations from Right Handed Neutrinos}, Phys. Rev. D 87 (2013) 093006.
\newblock \href {http://arxiv.org/abs/1208.4607} {\path{arXiv:1208.4607}}, \href {http://dx.doi.org/10.1103/PhysRevD.87.093006} {\path{doi:10.1103/PhysRevD.87.093006}}.

\bibitem{Ghiglieri:2020ulj}
J.~Ghiglieri, M.~Laine, {Sterile neutrino dark matter via coinciding resonances}, JCAP 07 (2020) 012.
\newblock \href {http://arxiv.org/abs/2004.10766} {\path{arXiv:2004.10766}}, \href {http://dx.doi.org/10.1088/1475-7516/2020/07/012} {\path{doi:10.1088/1475-7516/2020/07/012}}.

\bibitem{Khalil:2008kp}
S.~Khalil, O.~Seto, {Sterile neutrino dark matter in B - L extension of the standard model and galactic 511-keV line}, JCAP 10 (2008) 024.
\newblock \href {http://arxiv.org/abs/0804.0336} {\path{arXiv:0804.0336}}, \href {http://dx.doi.org/10.1088/1475-7516/2008/10/024} {\path{doi:10.1088/1475-7516/2008/10/024}}.

\bibitem{Abada:2014zra}
A.~Abada, G.~Arcadi, M.~Lucente, {Dark Matter in the minimal Inverse Seesaw mechanism}, JCAP 10 (2014) 001.
\newblock \href {http://arxiv.org/abs/1406.6556} {\path{arXiv:1406.6556}}, \href {http://dx.doi.org/10.1088/1475-7516/2014/10/001} {\path{doi:10.1088/1475-7516/2014/10/001}}.

\bibitem{Seto:2020udg}
O.~Seto, T.~Shimomura, {Signal from sterile neutrino dark matter in extra $U(1)$ model at direct detection experiment}, Phys. Lett. B 811 (2020) 135880.
\newblock \href {http://arxiv.org/abs/2007.14605} {\path{arXiv:2007.14605}}, \href {http://dx.doi.org/10.1016/j.physletb.2020.135880} {\path{doi:10.1016/j.physletb.2020.135880}}.

\bibitem{Abada:2014vea}
A.~Abada, M.~Lucente, {Looking for the minimal inverse seesaw realisation}, Nucl. Phys. B 885 (2014) 651--678.
\newblock \href {http://arxiv.org/abs/1401.1507} {\path{arXiv:1401.1507}}, \href {http://dx.doi.org/10.1016/j.nuclphysb.2014.06.003} {\path{doi:10.1016/j.nuclphysb.2014.06.003}}.

\bibitem{Casas:2001sr}
J.~A. Casas, A.~Ibarra, {Oscillating neutrinos and $\mu \to e, \gamma$}, Nucl. Phys. B 618 (2001) 171--204.
\newblock \href {http://arxiv.org/abs/hep-ph/0103065} {\path{arXiv:hep-ph/0103065}}, \href {http://dx.doi.org/10.1016/S0550-3213(01)00475-8} {\path{doi:10.1016/S0550-3213(01)00475-8}}.

\bibitem{Zyla:2020zbs}
P.~A. Zyla, et~al., {Review of Particle Physics}, PTEP 2020~(8) (2020) 083C01.
\newblock \href {http://dx.doi.org/10.1093/ptep/ptaa104} {\path{doi:10.1093/ptep/ptaa104}}.

\bibitem{Asaka:2006nq}
T.~Asaka, M.~Laine, M.~Shaposhnikov, {Lightest sterile neutrino abundance within the nuMSM}, JHEP 01 (2007) 091, [Erratum: JHEP 02, 028 (2015)].
\newblock \href {http://arxiv.org/abs/hep-ph/0612182} {\path{arXiv:hep-ph/0612182}}, \href {http://dx.doi.org/10.1088/1126-6708/2007/01/091} {\path{doi:10.1088/1126-6708/2007/01/091}}.

\bibitem{Boyarsky:2006jm}
A.~Boyarsky, A.~Neronov, O.~Ruchayskiy, M.~Shaposhnikov, {The Masses of active neutrinos in the nuMSM from X-ray astronomy}, JETP Lett. 83 (2006) 133--135.
\newblock \href {http://arxiv.org/abs/hep-ph/0601098} {\path{arXiv:hep-ph/0601098}}, \href {http://dx.doi.org/10.1134/S0021364006040011} {\path{doi:10.1134/S0021364006040011}}.

\bibitem{Hall:2009bx}
L.~J. Hall, K.~Jedamzik, J.~March-Russell, S.~M. West, {Freeze-In Production of FIMP Dark Matter}, JHEP 03 (2010) 080.
\newblock \href {http://arxiv.org/abs/0911.1120} {\path{arXiv:0911.1120}}, \href {http://dx.doi.org/10.1007/JHEP03(2010)080} {\path{doi:10.1007/JHEP03(2010)080}}.

\bibitem{DOnofrio:2014rug}
M.~D'Onofrio, K.~Rummukainen, A.~Tranberg, {Sphaleron Rate in the Minimal Standard Model}, Phys. Rev. Lett. 113~(14) (2014) 141602.
\newblock \href {http://arxiv.org/abs/1404.3565} {\path{arXiv:1404.3565}}, \href {http://dx.doi.org/10.1103/PhysRevLett.113.141602} {\path{doi:10.1103/PhysRevLett.113.141602}}.

\bibitem{Lello:2016rvl}
L.~Lello, D.~Boyanovsky, R.~D. Pisarski, {Production of heavy sterile neutrinos from vector boson decay at electroweak temperatures}, Phys. Rev. D 95~(4) (2017) 043524.
\newblock \href {http://arxiv.org/abs/1609.07647} {\path{arXiv:1609.07647}}, \href {http://dx.doi.org/10.1103/PhysRevD.95.043524} {\path{doi:10.1103/PhysRevD.95.043524}}.

\bibitem{Bezrukov:2014nza}
F.~Bezrukov, D.~Gorbunov, {Relic Gravity Waves and 7 keV Dark Matter from a GeV scale inflaton}, Phys. Lett. B 736 (2014) 494--498.
\newblock \href {http://arxiv.org/abs/1403.4638} {\path{arXiv:1403.4638}}, \href {http://dx.doi.org/10.1016/j.physletb.2014.07.060} {\path{doi:10.1016/j.physletb.2014.07.060}}.

\bibitem{Shaposhnikov:2006xi}
M.~Shaposhnikov, I.~Tkachev, {The nuMSM, inflation, and dark matter}, Phys. Lett. B 639 (2006) 414--417.
\newblock \href {http://arxiv.org/abs/hep-ph/0604236} {\path{arXiv:hep-ph/0604236}}, \href {http://dx.doi.org/10.1016/j.physletb.2006.06.063} {\path{doi:10.1016/j.physletb.2006.06.063}}.

\bibitem{Merle:2015oja}
A.~Merle, M.~Totzauer, {keV Sterile Neutrino Dark Matter from Singlet Scalar Decays: Basic Concepts and Subtle Features}, JCAP 06 (2015) 011.
\newblock \href {http://arxiv.org/abs/1502.01011} {\path{arXiv:1502.01011}}, \href {http://dx.doi.org/10.1088/1475-7516/2015/06/011} {\path{doi:10.1088/1475-7516/2015/06/011}}.

\bibitem{Heeck:2017xbu}
J.~Heeck, D.~Teresi, {Cold keV dark matter from decays and scatterings}, Phys. Rev. D 96~(3) (2017) 035018.
\newblock \href {http://arxiv.org/abs/1706.09909} {\path{arXiv:1706.09909}}, \href {http://dx.doi.org/10.1103/PhysRevD.96.035018} {\path{doi:10.1103/PhysRevD.96.035018}}.

\bibitem{Esteban:2020cvm}
I.~Esteban, M.~C. Gonzalez-Garcia, M.~Maltoni, T.~Schwetz, A.~Zhou, {The fate of hints: updated global analysis of three-flavor neutrino oscillations}, JHEP 09 (2020) 178.
\newblock \href {http://arxiv.org/abs/2007.14792} {\path{arXiv:2007.14792}}, \href {http://dx.doi.org/10.1007/JHEP09(2020)178} {\path{doi:10.1007/JHEP09(2020)178}}.

\bibitem{Das:2014jxa}
A.~Das, P.~S. Bhupal~Dev, N.~Okada, {Direct bounds on electroweak scale pseudo-Dirac neutrinos from $\sqrt s=8$ TeV LHC data}, Phys. Lett. B 735 (2014) 364--370.
\newblock \href {http://arxiv.org/abs/1405.0177} {\path{arXiv:1405.0177}}, \href {http://dx.doi.org/10.1016/j.physletb.2014.06.058} {\path{doi:10.1016/j.physletb.2014.06.058}}.

\bibitem{Das:2015toa}
A.~Das, N.~Okada, {Improved bounds on the heavy neutrino productions at the LHC}, Phys. Rev. D 93~(3) (2016) 033003.
\newblock \href {http://arxiv.org/abs/1510.04790} {\path{arXiv:1510.04790}}, \href {http://dx.doi.org/10.1103/PhysRevD.93.033003} {\path{doi:10.1103/PhysRevD.93.033003}}.

\bibitem{Klinger:2014vdo}
J.~Klinger, {Search for heavy Majorana neutrinos in $pp$ collisions at $\sqrt {s}$ = 8 TeV with the ATLAS detector}, Ph.D. thesis, Manchester U. (2014).

\bibitem{Sirunyan:2018mtv}
A.~M. Sirunyan, et~al., {Search for heavy neutral leptons in events with three charged leptons in proton-proton collisions at $\sqrt{s} =$ 13 TeV}, Phys. Rev. Lett. 120~(22) (2018) 221801.
\newblock \href {http://arxiv.org/abs/1802.02965} {\path{arXiv:1802.02965}}, \href {http://dx.doi.org/10.1103/PhysRevLett.120.221801} {\path{doi:10.1103/PhysRevLett.120.221801}}.

\bibitem{Fernandez-Martinez:2016lgt}
E.~Fernandez-Martinez, J.~Hernandez-Garcia, J.~Lopez-Pavon, {Global constraints on heavy neutrino mixing}, JHEP 08 (2016) 033.
\newblock \href {http://arxiv.org/abs/1605.08774} {\path{arXiv:1605.08774}}, \href {http://dx.doi.org/10.1007/JHEP08(2016)033} {\path{doi:10.1007/JHEP08(2016)033}}.

\bibitem{Shirai:2017jyz}
J.~Shirai, {Results and future plans for the KamLAND-Zen experiment}, J. Phys. Conf. Ser. 888~(1) (2017) 012031.
\newblock \href {http://dx.doi.org/10.1088/1742-6596/888/1/012031} {\path{doi:10.1088/1742-6596/888/1/012031}}.

\bibitem{Chanowitz:1978mv}
M.~S. Chanowitz, M.~A. Furman, I.~Hinchliffe, {Weak Interactions of Ultraheavy Fermions. 2.}, Nucl. Phys. B 153 (1979) 402--430.
\newblock \href {http://dx.doi.org/10.1016/0550-3213(79)90606-0} {\path{doi:10.1016/0550-3213(79)90606-0}}.

\bibitem{Durand:1989zs}
L.~Durand, J.~M. Johnson, J.~L. Lopez, {Perturbative Unitarity Revisited: A New Upper Bound on the Higgs Boson Mass}, Phys. Rev. Lett. 64 (1990) 1215.
\newblock \href {http://dx.doi.org/10.1103/PhysRevLett.64.1215} {\path{doi:10.1103/PhysRevLett.64.1215}}.

\bibitem{Korner:1992an}
J.~G. Korner, A.~Pilaftsis, K.~Schilcher, {Leptonic flavor changing Z0 decays in SU(2) x U(1) theories with right-handed neutrinos}, Phys. Lett. B 300 (1993) 381--386.
\newblock \href {http://arxiv.org/abs/hep-ph/9301290} {\path{arXiv:hep-ph/9301290}}, \href {http://dx.doi.org/10.1016/0370-2693(93)91350-V} {\path{doi:10.1016/0370-2693(93)91350-V}}.

\bibitem{Bernabeu:1993up}
J.~Bernabeu, J.~G. Korner, A.~Pilaftsis, K.~Schilcher, {Universality breaking effects in leptonic Z decays}, Phys. Rev. Lett. 71 (1993) 2695--2698.
\newblock \href {http://arxiv.org/abs/hep-ph/9307295} {\path{arXiv:hep-ph/9307295}}, \href {http://dx.doi.org/10.1103/PhysRevLett.71.2695} {\path{doi:10.1103/PhysRevLett.71.2695}}.

\bibitem{Fajfer:1998px}
S.~Fajfer, A.~Ilakovac, {Lepton flavor violation in light hadron decays}, Phys. Rev. D 57 (1998) 4219--4235.
\newblock \href {http://dx.doi.org/10.1103/PhysRevD.57.4219} {\path{doi:10.1103/PhysRevD.57.4219}}.

\bibitem{Ilakovac:1999md}
A.~Ilakovac, {Lepton flavor violation in the standard model extended by heavy singlet Dirac neutrinos}, Phys. Rev. D 62 (2000) 036010.
\newblock \href {http://arxiv.org/abs/hep-ph/9910213} {\path{arXiv:hep-ph/9910213}}, \href {http://dx.doi.org/10.1103/PhysRevD.62.036010} {\path{doi:10.1103/PhysRevD.62.036010}}.

\bibitem{Akhmedov:1998qx}
E.~K. Akhmedov, V.~A. Rubakov, A.~Y. Smirnov, {Baryogenesis via neutrino oscillations}, Phys. Rev. Lett. 81 (1998) 1359--1362.
\newblock \href {http://arxiv.org/abs/hep-ph/9803255} {\path{arXiv:hep-ph/9803255}}, \href {http://dx.doi.org/10.1103/PhysRevLett.81.1359} {\path{doi:10.1103/PhysRevLett.81.1359}}.

\bibitem{Bulbul:2014sua}
E.~Bulbul, M.~Markevitch, A.~Foster, R.~K. Smith, M.~Loewenstein, S.~W. Randall, {Detection of An Unidentified Emission Line in the Stacked X-ray spectrum of Galaxy Clusters}, Astrophys. J. 789 (2014) 13.
\newblock \href {http://arxiv.org/abs/1402.2301} {\path{arXiv:1402.2301}}, \href {http://dx.doi.org/10.1088/0004-637X/789/1/13} {\path{doi:10.1088/0004-637X/789/1/13}}.

\bibitem{Boyarsky:2014jta}
A.~Boyarsky, O.~Ruchayskiy, D.~Iakubovskyi, J.~Franse, {Unidentified Line in X-Ray Spectra of the Andromeda Galaxy and Perseus Galaxy Cluster}, Phys. Rev. Lett. 113 (2014) 251301.
\newblock \href {http://arxiv.org/abs/1402.4119} {\path{arXiv:1402.4119}}, \href {http://dx.doi.org/10.1103/PhysRevLett.113.251301} {\path{doi:10.1103/PhysRevLett.113.251301}}.

\bibitem{Fernandez-Martinez:2015hxa}
E.~Fernandez-Martinez, J.~Hernandez-Garcia, J.~Lopez-Pavon, M.~Lucente, {Loop level constraints on Seesaw neutrino mixing}, JHEP 10 (2015) 130.
\newblock \href {http://arxiv.org/abs/1508.03051} {\path{arXiv:1508.03051}}, \href {http://dx.doi.org/10.1007/JHEP10(2015)130} {\path{doi:10.1007/JHEP10(2015)130}}.

\bibitem{tHooft:1979rat}
G.~'t~Hooft, {Naturalness, chiral symmetry, and spontaneous chiral symmetry breaking}.

\bibitem{Gluza:2002vs}
J.~Gluza, {On teraelectronvolt Majorana neutrinos}, Acta Phys. Polon. B 33 (2002) 1735--1746.
\newblock \href {http://arxiv.org/abs/hep-ph/0201002} {\path{arXiv:hep-ph/0201002}}.

\bibitem{Shaposhnikov:2006nn}
M.~Shaposhnikov, {A Possible symmetry of the nuMSM}, Nucl. Phys. B 763 (2007) 49--59.
\newblock \href {http://arxiv.org/abs/hep-ph/0605047} {\path{arXiv:hep-ph/0605047}}, \href {http://dx.doi.org/10.1016/j.nuclphysb.2006.11.003} {\path{doi:10.1016/j.nuclphysb.2006.11.003}}.

\bibitem{Kersten:2007vk}
J.~Kersten, A.~Y. Smirnov, {Right-Handed Neutrinos at CERN LHC and the Mechanism of Neutrino Mass Generation}, Phys. Rev. D 76 (2007) 073005.
\newblock \href {http://arxiv.org/abs/0705.3221} {\path{arXiv:0705.3221}}, \href {http://dx.doi.org/10.1103/PhysRevD.76.073005} {\path{doi:10.1103/PhysRevD.76.073005}}.

\bibitem{Abada:2007ux}
A.~Abada, C.~Biggio, F.~Bonnet, M.~B. Gavela, T.~Hambye, {Low energy effects of neutrino masses}, JHEP 12 (2007) 061.
\newblock \href {http://arxiv.org/abs/0707.4058} {\path{arXiv:0707.4058}}, \href {http://dx.doi.org/10.1088/1126-6708/2007/12/061} {\path{doi:10.1088/1126-6708/2007/12/061}}.

\bibitem{Moffat:2017feq}
K.~Moffat, S.~Pascoli, C.~Weiland, {Equivalence between massless neutrinos and lepton number conservation in fermionic singlet extensions of the Standard Model}\href {http://arxiv.org/abs/1712.07611} {\path{arXiv:1712.07611}}.

\bibitem{Lopez-Pavon:2015cga}
J.~Lopez-Pavon, E.~Molinaro, S.~T. Petcov, {Radiative Corrections to Light Neutrino Masses in Low Scale Type I Seesaw Scenarios and Neutrinoless Double Beta Decay}, JHEP 11 (2015) 030.
\newblock \href {http://arxiv.org/abs/1506.05296} {\path{arXiv:1506.05296}}, \href {http://dx.doi.org/10.1007/JHEP11(2015)030} {\path{doi:10.1007/JHEP11(2015)030}}.

\bibitem{Abada:2015rta}
A.~Abada, G.~Arcadi, V.~Domcke, M.~Lucente, {Lepton number violation as a key to low-scale leptogenesis}, JCAP 11 (2015) 041.
\newblock \href {http://arxiv.org/abs/1507.06215} {\path{arXiv:1507.06215}}, \href {http://dx.doi.org/10.1088/1475-7516/2015/11/041} {\path{doi:10.1088/1475-7516/2015/11/041}}.

\bibitem{Abada:2017ieq}
A.~Abada, G.~Arcadi, V.~Domcke, M.~Lucente, {Neutrino masses, leptogenesis and dark matter from small lepton number violation?}, JCAP 12 (2017) 024.
\newblock \href {http://arxiv.org/abs/1709.00415} {\path{arXiv:1709.00415}}, \href {http://dx.doi.org/10.1088/1475-7516/2017/12/024} {\path{doi:10.1088/1475-7516/2017/12/024}}.

\bibitem{Abada:2018oly}
A.~Abada, G.~Arcadi, V.~Domcke, M.~Drewes, J.~Klaric, M.~Lucente, {Low-scale leptogenesis with three heavy neutrinos}, JHEP 01 (2019) 164.
\newblock \href {http://arxiv.org/abs/1810.12463} {\path{arXiv:1810.12463}}, \href {http://dx.doi.org/10.1007/JHEP01(2019)164} {\path{doi:10.1007/JHEP01(2019)164}}.

\end{thebibliography}

\end{document}